\documentclass[published]{nst}

\usepackage{subfigure,dcolumn}
\usepackage{epstopdf}
\usepackage{mhchem}
\usepackage{upgreek}

\begin{document}

\title{Charge resolution in the Isochronous Mass Spectrometry and the mass of $^{51}$Co}
\thanks{This work is supported by the National Key R\&D Program of China (Grants No. 2016YFA0400504, No. 2018YFA0404401),
	the Strategic Priority Research Program of Chinese Academy of Sciences (Grant No. XDB34000000),
	the NSFC (Grants No. 11905261, No. 11805032, No. 11975280, No. 11605248),
	Y.M.Xing thanks for support from CAS "Light of West China" Program,
	X.X. acknowledge support from China Postdoctoral Science Foundation (Contract NO. 2019M660250).
	X.Y. acknowledge partial support by the FRIB-CSC Fellow-ship, China under Grant No. 201704910964,
	R.J. Chen is supported by the International Postdoctoral Exchange Fellowship Program 2017 by the Office of China Postdoctoral Council (No.60 Document of OCPC, 2017),
	Y.A.L. receives support from the European Research Council (ERC) under the European Union's Horizon 2020 research and innovation programme (grant agreement No 682841 ``ASTRUm'').
}

\author{Xu Zhou}
\affiliation{Institute of Modern Physics, Chinese Academy of Sciences, Lanzhou 730000, China}
\affiliation{School of Nuclear Science and Technology, University of Chinese Academy of Sciences, Beijing 100049, People's Republic of China}

\author{Meng Wang}
\email[Corresponding author, ]{wangm@impcas.ac.cn}
\affiliation{Institute of Modern Physics, Chinese Academy of Sciences, Lanzhou 730000, China}
\affiliation{School of Nuclear Science and Technology, University of Chinese Academy of Sciences, Beijing 100049, People's Republic of China}

\author{Yu-Hu Zhang}
\email[Corresponding author, ]{yhzhang@impcas.ac.cn}
\affiliation{Institute of Modern Physics, Chinese Academy of Sciences, Lanzhou 730000, China}
\affiliation{School of Nuclear Science and Technology, University of Chinese Academy of Sciences, Beijing 100049, People's Republic of China}

\author{Hu-Shan Xu}
\affiliation{Institute of Modern Physics, Chinese Academy of Sciences, Lanzhou 730000, China}
\affiliation{School of Nuclear Science and Technology, University of Chinese Academy of Sciences, Beijing 100049, People's Republic of China}

\author{You-Jin Yuan}
\affiliation{Institute of Modern Physics, Chinese Academy of Sciences, Lanzhou 730000, China}
\affiliation{School of Nuclear Science and Technology, University of Chinese Academy of Sciences, Beijing 100049, People's Republic of China}

\author{Jian-Cheng Yang}
\affiliation{Institute of Modern Physics, Chinese Academy of Sciences, Lanzhou 730000, China}
\affiliation{School of Nuclear Science and Technology, University of Chinese Academy of Sciences, Beijing 100049, People's Republic of China}

\author{Yu. A. Litvinov}
\affiliation{Institute of Modern Physics, Chinese Academy of Sciences, Lanzhou 730000, China}
\affiliation{GSI Helmholtzzentrum f{\"u}r Schwerionenforschung, Planckstra{\ss }e 1, 64291 Darmstadt, Germany}

\author{S. A. Litvinov}
\affiliation{Institute of Modern Physics, Chinese Academy of Sciences, Lanzhou 730000, China}
\affiliation{GSI Helmholtzzentrum f{\"u}r Schwerionenforschung, Planckstra{\ss }e 1, 64291 Darmstadt, Germany}

\author{Bo Mei}
\affiliation{Institute of Modern Physics, Chinese Academy of Sciences, Lanzhou 730000, China}
\affiliation{Sino-French Institute of Nuclear Engineering and Technology, Sun Yat-sen University, Zhuhai 519082, China}

\author{Xin-Liang Yan}
\affiliation{Institute of Modern Physics, Chinese Academy of Sciences, Lanzhou 730000, China}
\affiliation{Argonne National Laboratory, 9700 S. Cass Avenue, Lemont, IL 60439, USA}

\author{Xing Xu}
\affiliation{Institute of Modern Physics, Chinese Academy of Sciences, Lanzhou 730000, China}
\affiliation{School of Science, Xi'an Jiaotong University, Xi'an 710049, China}

\author{Peng Shuai}
\affiliation{Institute of Modern Physics, Chinese Academy of Sciences, Lanzhou 730000, China}

\author{Yuan-Ming Xing}
\affiliation{Institute of Modern Physics, Chinese Academy of Sciences, Lanzhou 730000, China}

\author{Rui-Jiu Chen}
\affiliation{Institute of Modern Physics, Chinese Academy of Sciences, Lanzhou 730000, China}
\affiliation{GSI Helmholtzzentrum f{\"u}r Schwerionenforschung, Planckstra{\ss }e 1, 64291 Darmstadt, Germany}

\author{Xiang-Cheng Chen}
\affiliation{Institute of Modern Physics, Chinese Academy of Sciences, Lanzhou 730000, China}

\author{Chao-Yi Fu}
\affiliation{Institute of Modern Physics, Chinese Academy of Sciences, Lanzhou 730000, China}

\author{Qi Zeng}
\affiliation{Institute of Modern Physics, Chinese Academy of Sciences, Lanzhou 730000, China}
\affiliation{School of nuclear science and engineering, East China University of Technology, Nanchang 330013, China}

\author{Ming-Ze Sun}
\affiliation{Institute of Modern Physics, Chinese Academy of Sciences, Lanzhou 730000, China}

\author{Hong-Fu Li}
\affiliation{Institute of Modern Physics, Chinese Academy of Sciences, Lanzhou 730000, China}
\affiliation{School of Nuclear Science and Technology, University of Chinese Academy of Sciences, Beijing 100049, People's Republic of China}

\author{Qian Wang}
\affiliation{Institute of Modern Physics, Chinese Academy of Sciences, Lanzhou 730000, China}
\affiliation{School of Nuclear Science and Technology, University of Chinese Academy of Sciences, Beijing 100049, People's Republic of China}

\author{Tong Bao}
\affiliation{Institute of Modern Physics, Chinese Academy of Sciences, Lanzhou 730000, China}

\author{Min Zhang}
\affiliation{Institute of Modern Physics, Chinese Academy of Sciences, Lanzhou 730000, China}
\affiliation{School of Nuclear Science and Technology, University of Chinese Academy of Sciences, Beijing 100049, People's Republic of China}

\author{Min Si}
\affiliation{Institute of Modern Physics, Chinese Academy of Sciences, Lanzhou 730000, China}
\affiliation{Universit{\'e} Paris-Saclay, CNRS/IN2P3, IJCLab, F-91405 Orsay, France}

\author{Han-Yu Deng}
\affiliation{Institute of Modern Physics, Chinese Academy of Sciences, Lanzhou 730000, China}
\affiliation{School of Nuclear Science and Technology, University of Chinese Academy of Sciences, Beijing 100049, People's Republic of China}

\author{Ming-Zheng Liu}
\affiliation{Institute of Modern Physics, Chinese Academy of Sciences, Lanzhou 730000, China}
\affiliation{School of Nuclear Science and Technology, University of Chinese Academy of Sciences, Beijing 100049, People's Republic of China}

\author{Ting Liao}
\affiliation{Institute of Modern Physics, Chinese Academy of Sciences, Lanzhou 730000, China}
\affiliation{School of Nuclear Science and Technology, University of Chinese Academy of Sciences, Beijing 100049, People's Republic of China}

\author{Jin-Yang Shi}
\affiliation{Institute of Modern Physics, Chinese Academy of Sciences, Lanzhou 730000, China}
\affiliation{School of Nuclear Science and Technology, University of Chinese Academy of Sciences, Beijing 100049, People's Republic of China}

\author{Yu-Nan Song}
\affiliation{Institute of Modern Physics, Chinese Academy of Sciences, Lanzhou 730000, China}
\affiliation{School of Nuclear Science and Technology, University of Chinese Academy of Sciences, Beijing 100049, People's Republic of China}

\begin{abstract}
Isochronous mass spectrometry (IMS) at heavy-ion storage rings is a powerful tool for mass measurements of short-lived nuclei.
In IMS experiments, masses are determined through precision measurements of the revolution times of the ions stored in the ring.
However, the revolution times can not be resolved for the particles with nearly the same mass-to-charge ($m/q$) ratios. 
In order to overcome this limitation and to extract the accurate revolution times for such pairs of ion species with very close $m/q$ ratios, the amplitudes of timing signals from the detector, based on the emission of secondary electrons, have been analyzed in our early work on the particle identification. 
Here, the previous data analysis method is further improved by taking into account the signal amplitudes, the detection efficiencies, as well as the number of stored ions in the ring. 
A sensitive $Z$-dependent parameter is introduced in the data analysis, leading to a better resolution of $^{34}$Ar$^{18+}$ and $^{51}$Co$^{27+}$ with $A/Z=17/9$. 
The mean revolution times of $^{34}$Ar$^{18+}$ and $^{51}$Co$^{27+}$ are deduced although their time difference is merely 1.8~ps. 
The uncorrected, overlapped peak of these ions has FWHM=7.7 ps. 
The mass excess of $^{51}$Co is re-determined to be $-27332(41)$ keV in agreement with the previous value of $-27342(48)$ keV.
\end{abstract}


\maketitle

\section{Introduction}

Isochronous mass spectrometry (IMS) at heavy-ion storage rings plays an important role in mass measurements of short-lived nuclei \cite{1,1_1}.
There are three storage ring facilities conducting such experiments worldwide \cite{2_1,2_2, 2,Steck}, namely the experimental cooler-storage ring (CSRe) in Lanzhou, the experimental storage ring (ESR) in Darmstadt and the Rare-RI Ring (R3) in Saitama.
In the IMS experiments, nuclei of interest are produced in projectile fragmentation (PF) or in-flight fission nuclear reactions,  where the production cross sections are related to the binding energies of the fragments \cite{3_1,3_2,3_3}.
The reaction products of interest are selected with a fragment separator and injected into the storage ring for mass measurements.
The revolution times of the stored nuclides are directly related to their mass-to-charge ($m/q$) ratios.
The revolution times are typically measured by using dedicated timing detectors \cite{8_1,8,TOF,8_2}.
It is obvious that only those revolution times which could unambiguously be assigned to a specific ion species can be used  \cite{9,10}.

Particle identification (PID) is a prerequisite in experiments with radioactive ion beams (RIB) produced in a nuclear reaction, 
and is routinely accomplished along a RIB beam-line by measuring the time-of-flight (TOF), energy loss ($\Delta$E), and magnetic rigidity (B$\rho$) for each particle \cite{Liang, Sun}.
In particular, the atomic number $Z$ is mainly determined by using the $\Delta$E value which is usually measured with various ionisation-type detectors placed along the beam-line.

Among the storage-ring based facilities, PID within a beam-line can only be done at the R3 in RIKEN \cite{r3,yamaguchi}, 
where the quasi-DC beam from the superconducting cyclotron is employed.
Thus, the RIBs can indeed be identified on the event-by-event basis prior to their injection into the R3 .
It is very different in the cases of the CSRe and ESR.
The driver accelerator at the facility in Lanzhou is the heavy-ion synchrotron CSRm \cite{CSRm}.
The primary beams are fast extracted from the CSRm within $\sim$300~ns.
Correspondingly the produced RIBs have a similar bunched structure.
Therefore, the event-by-event PID is not feasible in the radioactive-ion beam line RIBLL2 connecting the CSRm and the CSRe.
Similar situation is at GSI in Darmstadt, where the ions are fast-extracted from the heavy-ion synchrotron SIS-18, analyzed in flight with the fragment separator FRS 
and injected for isochronous mass measurements into the experimental storage ring ESR \cite{1_5}.
In this work we examine the sensitivity of the timing detectors to the charge of the stored ions for the purpose of the in-ring PID.

By using IMS at the CSRe, masses of more than 30 short-lived nuclei have been measured \cite{Fu20,3,4,5,6,7,52Co,Xu15,9_1,add1,add2}.
For a recent review the reader is referred to Refs. \cite{2, Steck,7} and references cited therein.
In these experiments, the ions are identified almost solely through their $m/q$ ratios 
by comparing the revolution time spectrum obtained experimentally with the simulated one \cite{9,10}.
The $m/q$ resolving power of $\sim3\times 10^5$ has been achieved at the CSRe, 
which e.g. is sufficient to identify the isomeric states in $^{101}$In and $^{52}$Co \cite{4,52Co}.
However, for some pairs of ions with very similar $m/q$ ratios, such as $^{51}$Co$^{27+}$ and $^{34}$Ar$^{18+}$, 
the revolution times are too close to each other to be resolved clearly. 
The charge determination can provide crucial information for PID in these cases. The charge-resolved IMS was developed by analyzing the average amplitudes of signals from the timing detector induced by stored ions \cite{11}. This method was used successfully to determine for the first time the mass of $^{51}$Co.
Although the charge difference ($\Delta Z$) of $^{51}$Co$^{27+}$ and $^{34}$Ar$^{18+}$ is relatively large, some events could not be unambiguously identified due to the limited charge resolution in that work \cite{11}.

A higher charge resolution is desired. 
For example, masses of $^{54,56}$Ti nuclides have been measured in an earlier experiment at the CSRe \cite{3,Xu15}.
The $^{55}$Ti nuclide must have been produced with a significant amount in this experiment according to the yields of neighbouring nuclei, 
but it could not be separated from $^{50}$Ca.
To measure the mass of $^{55}$Ti, it is necessary to resolve $^{55}$Ti$^{22+}$ and $^{50}$Ca$^{20+}$ by charge, where $\Delta Z$ is merely 2. 
Such high charge resolution was not realized in that experiment.
Due to the limited charge resolution, mass measurements of $N=Z$ nuclides with conventional IMS at the CSRe or ESR are usually regarded to be unfeasible. 
The $N=Z$ nuclides with $\Delta Z=2$ have to be resolved by charge, while the neighbouring $N=Z$ ($\Delta Z=1$) nuclides 
can be resolved by their revolution times because the difference of their $m/q$ ratios is sufficiently large, thanks to the odd-even staggering of binding energies.
If ions with $\Delta Z=2$ can be resolved by charge then, in principle, all nuclides can be identified and measured in IMS experiments.

In this work, we describe a new method, in which detailed characteristics of the timing detector are considered in the data analysis. 
The charge resolution of IMS is improved significantly, leading to a complete separation of $^{34}$Ar and $^{51}$Co.
Possible future developments are discussed as well.

\section{Experiment and the timing detector}

The following discussions are based on the same experimental data that were analyzed in the earlier work \cite{11}.
More details about the experiment can be found in Refs. \cite{11,12,13}.
This experiment aimed to measure the masses of $T_z=-3/2$ nuclei in the $pf$-shell.
The nuclei of interest were produced by using projectile fragmentation of 463~MeV/u $^{58}$Ni$^{19+}$ primary beams in 
a $\sim 15$ mm $^9$Be production target.
At this energy the fragments emerged from the target predominantly as bare ions.
Therefore, in the following we assume the atomic charge $q$ to be equal to atomic number $Z$.
The RIBLL2 and CSRe were set to a fixed magnetic rigidity of $B\rho = 5.677$~Tm.
All fragments within the $B\rho$-acceptance of $\sim\pm$0.2$\%$ were injected into the CSRe.

\begin{figure}[!htb]\centering
	\includegraphics[angle=0,width=8cm]{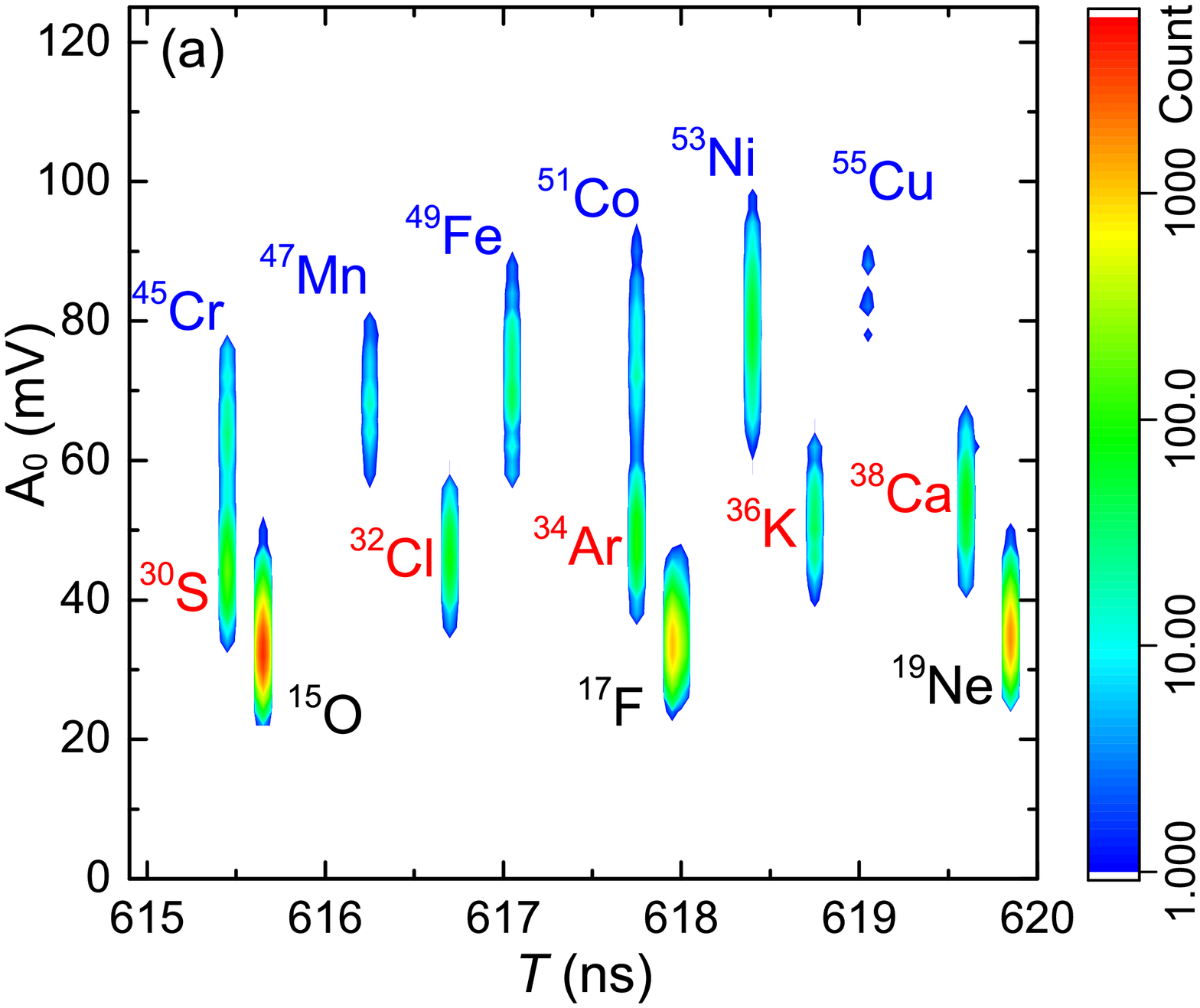}
	
	\vspace{2mm}
	\includegraphics[angle=0,width=8cm]{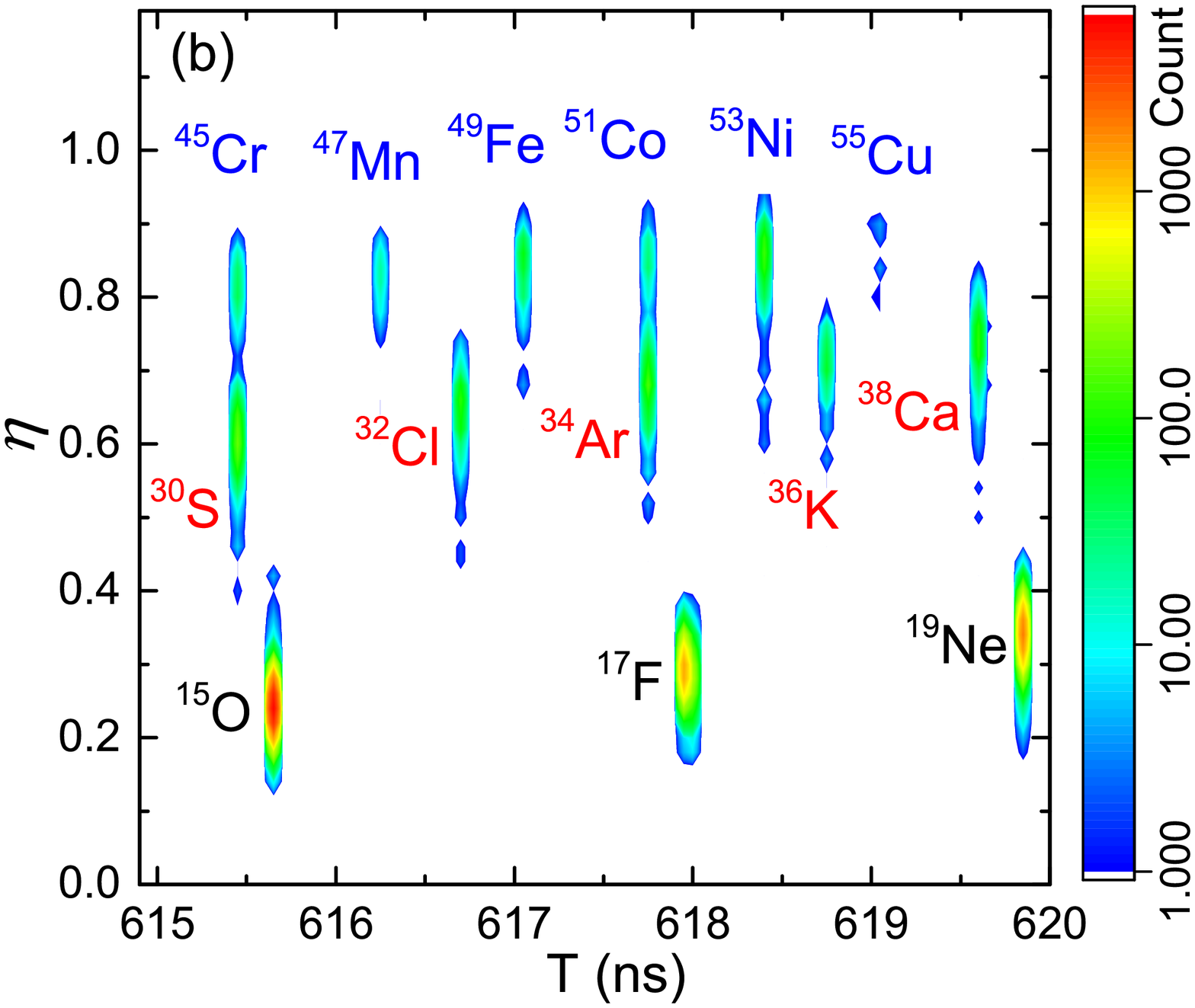}
	\caption{(Colour online)
		Two-dimensional plot of average signal amplitudes $A_0$ (a) and detection efficiencies $\eta$ (b) as a function of revolution time $T$, for $T_z=-1/2$ (black labels), $T_z=-1$ (red labels) and $T_z=-3/2$ (blue labels) ions.
		\label{fig1}}
\end{figure}
\begin{figure*}\centering
	\includegraphics[angle=0,width=5.5cm]{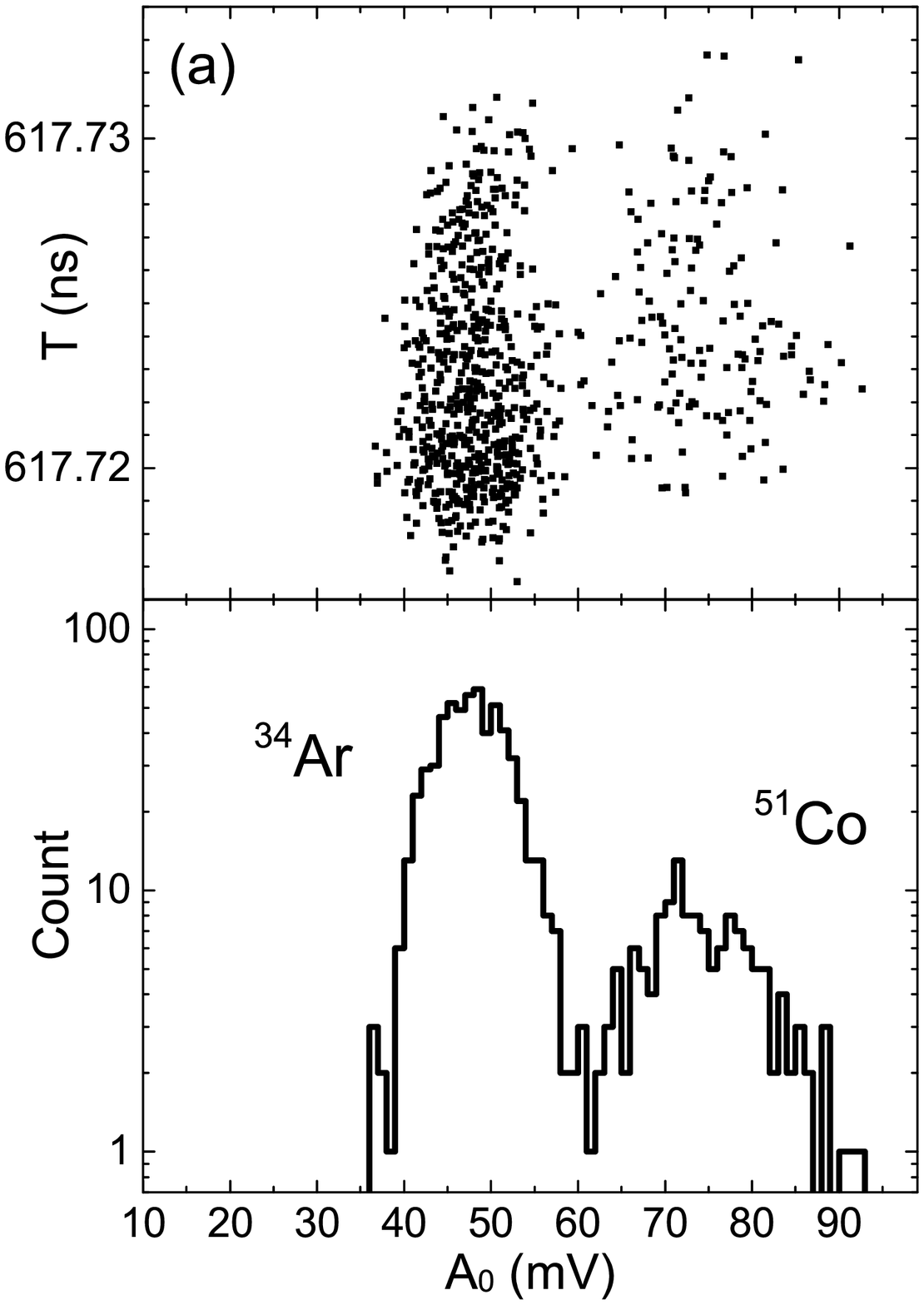}
	\includegraphics[angle=0,width=5.5cm]{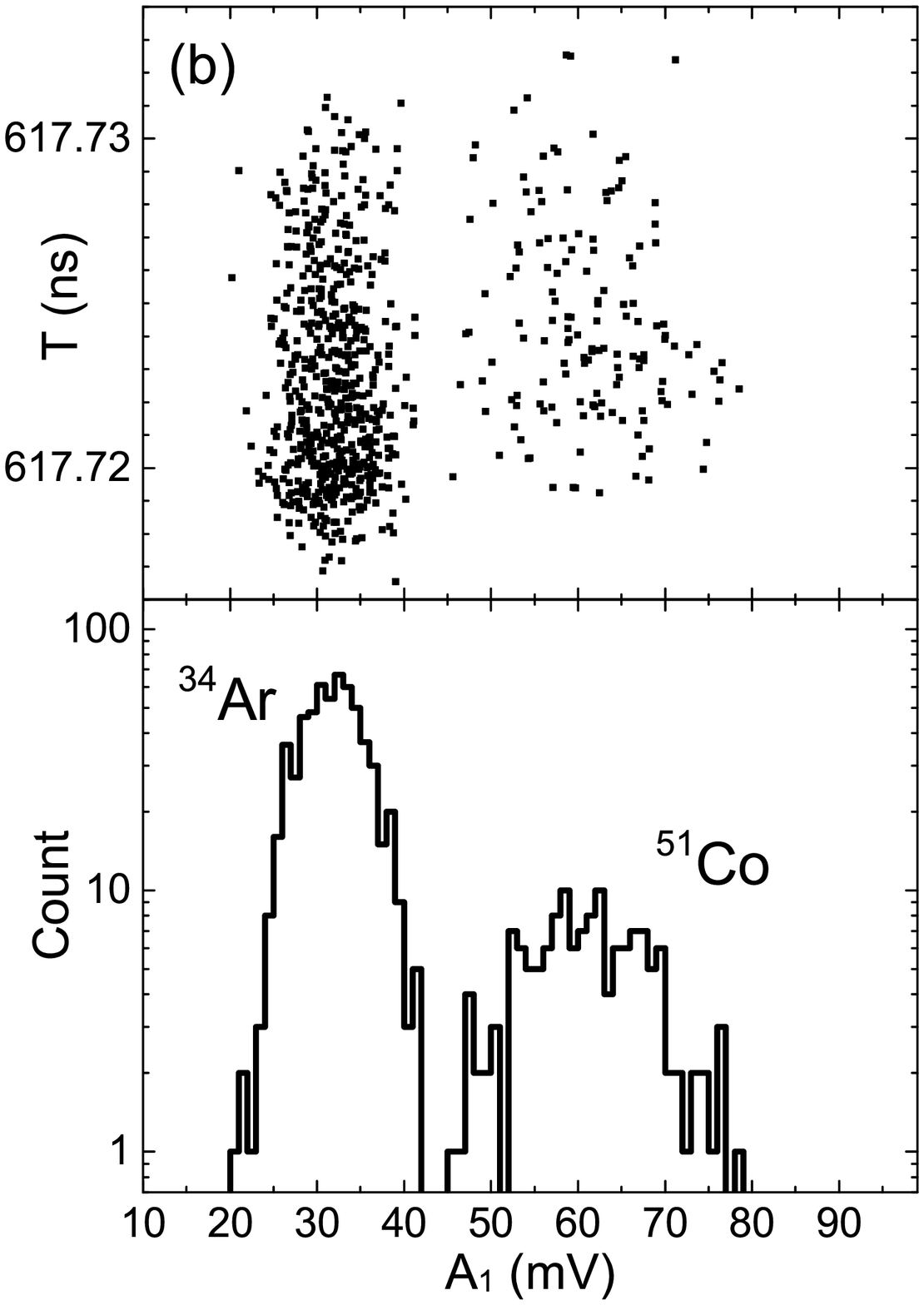}
	\includegraphics[angle=0,width=5.5cm]{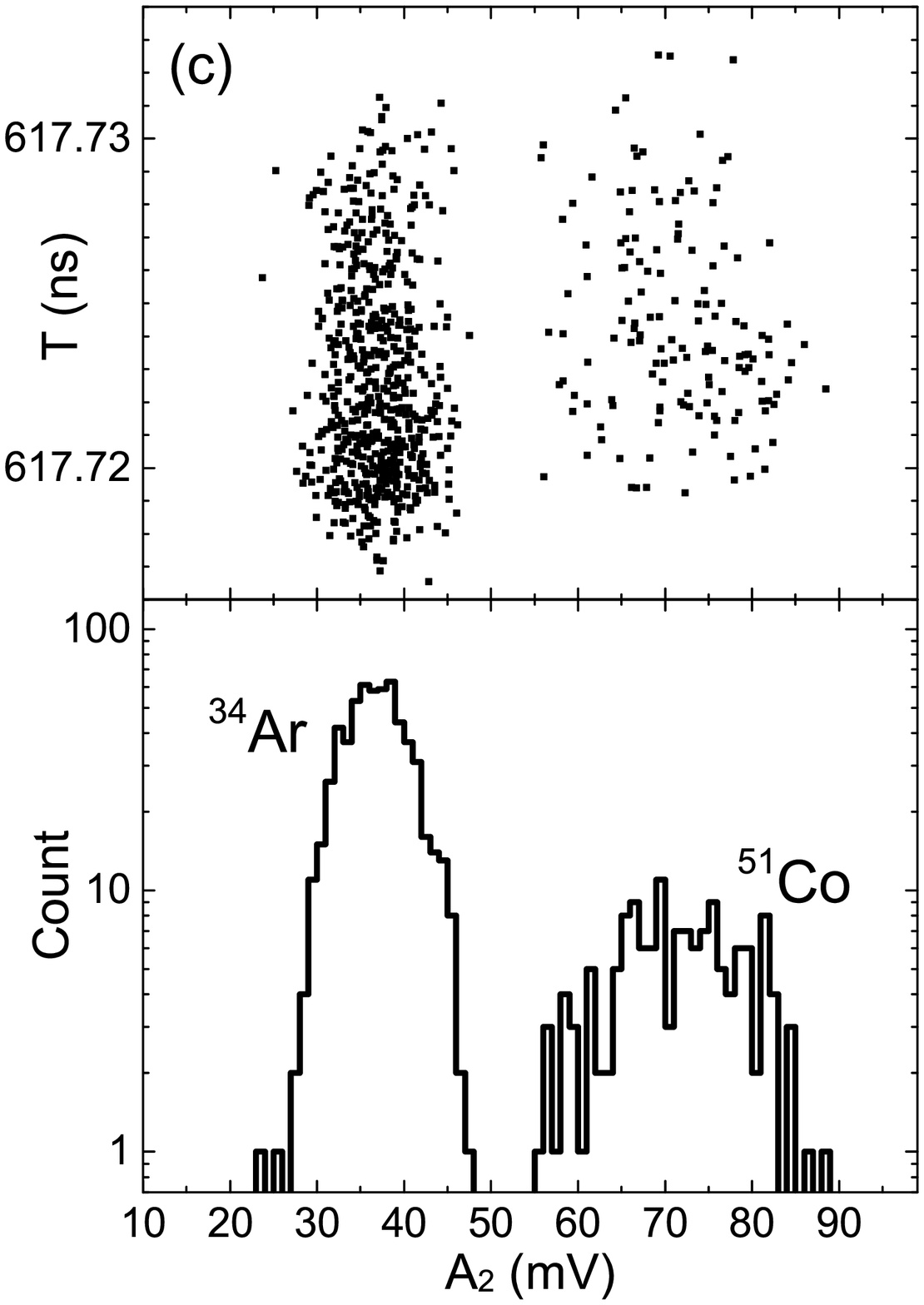}
	\caption{
		The scatter plot of the revolution times $T$ as a function of $A_0$ (a), $A_1$ (b), $A_2$ (c) with their projected histogram for $^{34}$Ar and $^{51}$Co ions.
		\label{fig2}}
\end{figure*}

The revolution times of the stored ions were measured by using a timing detector equipped 
with a carbon foil of $19~\mu \rm{g/cm^2}$ in thickness and $40$~mm in diameter \cite{8}.
The foil was positioned in the circulation path of the stored ions.
When an ion passed through the carbon foil, secondary electrons were released from the foil surface and guided isochronously to a set of micro-channel plates (MCP) in the Chevron configuration \cite{15}.
The MCP had the same diameter as the carbon foil and the micro-channels were 5 $\mu$m in diameter.
When an electron hits a channel of the MCP, an avalanche of further secondary electrons propagated through the channel.
The bunch of electrons exiting from the rear side of that channel was collected by a conical anode.
A signal can be generated with the activation of one or more channels.
The negative-voltage signal was directly transmitted from the anode to a digital oscilloscope for data acquisition.
The recording time was $200~\mu$s for each injection in this experiment.

The amount of electrons from the carbon foil induced by an ion is roughly proportional to the energy loss, which scales with $q^2=Z^{2}$ \cite{8}.
Consequently, both the average signal amplitude ($A_0$) and the detection efficiency ($\eta$) for an ion depend strongly on its atomic number $Z$.

Figure~\ref{fig1} (a) shows the two-dimensional plot of revolution times $T$ and $A_0$ obtained in the experiment, similar as Fig.~3 given in the earlier work~\cite{11}.
The $^{51}$Co$^{27+}$ and $^{34}$Ar$^{18+}$ ions can be resolved by using the $A_0$ value, albeit not completely separated.
Figure~\ref{fig1} (a) is an analogue of the typical PID plots at an in-flight separator \cite{Liang,Sun}, though with a lower charge resolution but a higher $m/q$ resolution. Figure~\ref{fig1} (b) shows the two-dimensional plot of $T$ versus $\eta$.
The two plots in Fig.~\ref{fig1} are quite similar. It seems that $A_0$ provides a better charge resolution for heavier ions while $\eta$ does better for lighter ones.

\section{Analysis of the signal amplitudes}

In the earlier work, the average signal amplitude ($A_0$) was used to resolve the $^{51}$Co$^{27+}$ and $^{34}$Ar$^{18+}$ ions \cite{11}. 
Figure~\ref{fig2} (a) shows the scatter plot of revolution times $T$ versus $A_0$ and its projected histogram for $^{34}$Ar and $^{51}$Co. 
Although two ion species can be well resolved, there still exists an overlap between the two Gaussian-like distributions.

Assuming each micro-channel is fully discharged after the electron impact and thus provides the same contribution to the signal, 
$A_0$ is roughly proportional to the average number of the activated micro-channels and in turn to the number of released secondary electrons from the foil surface.
However, there were ion passages through the detector which did not generate signals. This reflects the finite detection efficiency $\eta$ \cite{11},
which includes electron hits in between the channels (geometrical efficiency) and the hits into the already discharged channels (dynamic efficiency).
We define an average signal amplitude corrected by the total efficiency as $A_1$

\begin{equation}
A_1=A_0 \times \eta.
\label{eq1}
\end{equation}
It is clear that $A_1$ is proportional to the total number of micro-channels activated by an ion.
Figure~\ref{fig2}(b) shows the scatter plot of revolution times $T$ versus $A_1$ and the projected histogram for $^{34}$Ar and $^{51}$Co. 
It can directly be seen that the separation of these two ions became better.

After a micro-channel was activated to output the avalanche electrons, it took $\sim 10$~ms to recharge the channel.
During this period, the channel did not respond to other coming electrons.
Since the measurement time is much shorter than the recharging time, each micro-channel could be used at most once during the measurement~\cite{ps}.
If the number of stored ions was too large in one injection into the CSRe, 
the total number of activated micro-channels would be large and
the chance for each electron to activate a micro-channel would be reduced, leading to the reduced $A_0$ and $\eta$ values.
This dynamic efficiency obviously depends on the number of stored ions and their charge.
The sum amplitude of all signals ($A_s$) was a direct measure of the total number of activated channels in one injection. 
It is thus expected that $A_1$ values have a negative correlation with $A_s$ values for one ion species.

\begin{figure}[h]\centering
	\includegraphics[angle=0,width=8cm]{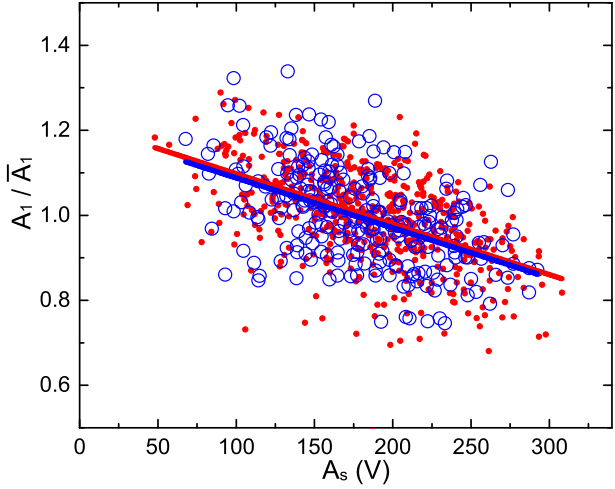}
	\caption{(Colour online)
		For $^{36}$K (blue empty circles) and $^{53}$Ni (red solid circles) ions, plot of the constructed parameter $A_1$ values divided by the mean value $\overline{A}_1$ for such ion specie, as a function of the sum amplitude of all signals in one injection ($A_{s}$).
		\label{fig3}}
\end{figure}

In the present experiment, the number of stored ions ranged from 2 to 46 in various injections with the most probable value of 21.
The $A_s$ values ranged from $50$ to $300$ V. 
The relationship between $A_1$ and $A_s$ values is similar for all ion species, and is demonstrated in Fig.~\ref{fig3} on the example of $^{36}$K and $^{53}$Ni ions, which can be clearly identified by using their revolution times. 
The mean $\bar{A}_1$ values are calculated for these two ion species. 
The $A_1/\bar{A}_1$ ratios are calculated for each individual ion and the scatter plot of $A_1/\bar{A}_1$ versus $A_s$ is shown in Fig.~\ref{fig3}. 
The $A_1/\bar{A}_1$ values for both ion species are described by a linear function of $A_s$.
The obtained two slopes are very similar indicating a constant slope for all ion species.
If the observed correlation of $A_1$ and $A_s$ values is taken into account, the charge resolution can be improved.

A new parameter $A_2$ is defined as an estimate of the $A_1$ value at $A_s = 0$~V.
The $A_2$ value is proportional to the number of channels that should be activated by the ion if all micro-channels are available.
It is thus directly related to the number of released electrons from the detector foil due to this ion.
Considering the constant slope for different ion species as shown in Fig.~\ref{fig3}, the $A_2$ value can be calculated as

\begin{equation}
A_2=A_1 \times (1-B\times A_s).
\label{eq3}
\end{equation}
The $B$ parameter represents the slope in Fig.~\ref{fig3} and is determined by using all ion species with proton number $Z>8$ and more than 100 counts. 
It is equal to $B=-9.05(7)\times 10^{-4} \rm{~V^{-1}}$.
Figure~\ref{fig2}(c) shows the scatter plot of revolution times $T$ versus $A_2$ and the projected histogram for $^{34}$Ar and $^{51}$Co ions.
It can be seen that the two ions are completely separated.

Since the revolution time $T$ is $B\rho$ dependent for ions of $^{34}$Ar and $^{51}$Co~\cite{2_1,2_2}, the $T$ values should depend on the ions' average passage position through the detector.
As show in Fig.~\ref{fig2}, there is no correlation of $A_0$, $A_1$ and $A_2$ values with the revolution time $T$, proving position independence of $A_0$, $A_1$ and $A_2$ in this experiment.

By using $A_2$ for particle identification, the mass excess of $^{51}$Co is re-determined to be $-27332(41)$ keV, in agreement with the value $-27342(48)$ keV obtained previously \cite{11}. 
The smaller mass uncertainty in this analysis is mainly due to larger statistics obtained to less strict constraints. 
In the earlier work only the particles that were stored for more than 300 revolutions in the CSRe were used in the analysis. 
Based on the since then improved data analysis methods, the threshold could be lowered in the present work to 80 revolutions, thus increasing the statistics.

\section{Charge resolution}

\begin{figure}[h]\centering
	\includegraphics[angle=0,width=8cm]{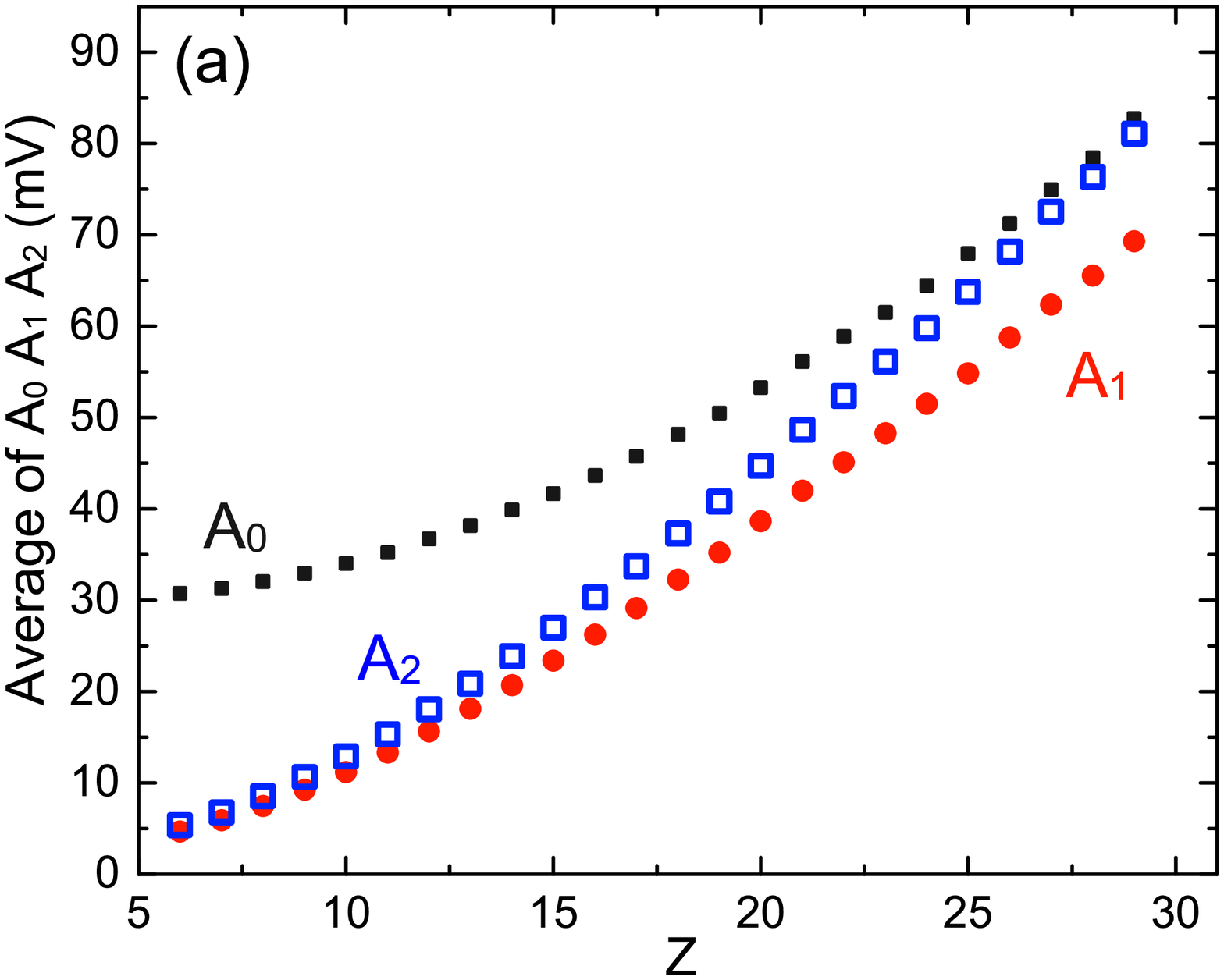}
	
	\vspace{2mm}
	\includegraphics[angle=0,width=8cm]{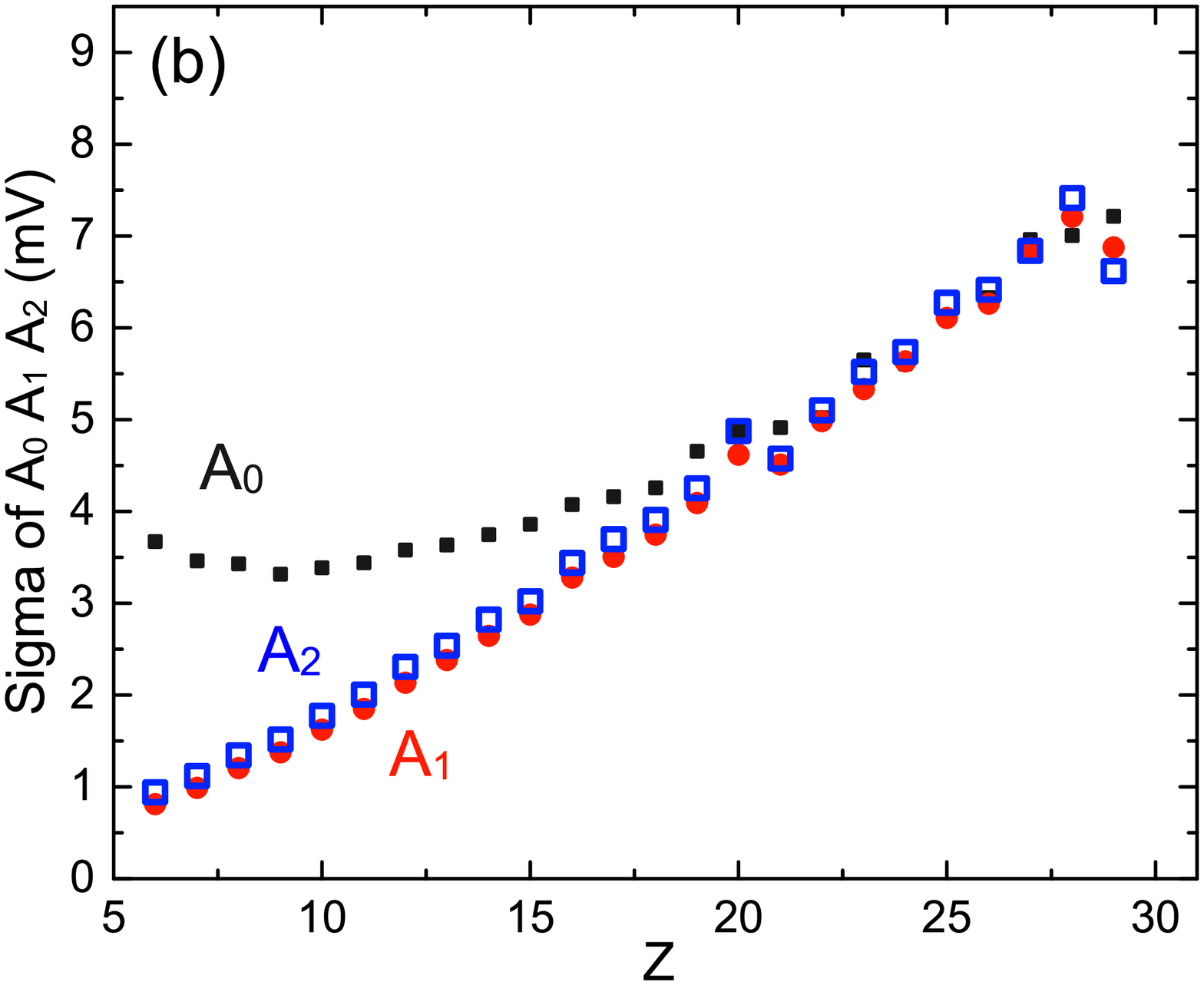}
	\caption{ (Colour online)
		For ions with same proton number $Z$, plot of average (a) and standard deviations (b) of $A_0$ (black solid squares), $A_1$ (red solid circles) and $A_2$ (blue empty squares) as a function of $Z$.
		\label{fig4}}
\end{figure}

The $A_0$, $A_1$ and $A_2$ values are calculated for all ions.
Figure~\ref{fig4} (a) shows the average values of these three parameters for ions with the same proton number $Z$.
As can be seen, the $A_1$ and $A_2$ values are very similar, while $A_0$ is larger than the other two parameters for light ions.
$A_2$ increases rapidly with proton number and approaches $A_0$.
It is evident that $A_2$ is more sensitive to proton number $Z$. 
The standard deviations of these three parameters are shown in Fig.~\ref{fig4}(b) as a function of $Z$.
For heavy ions ($Z\sim20$) they are similar for all three parameters, while for lighter ions the standard deviation of $A_0$ is considerably larger.

The charge resolution ratio $R$ is introduced to represent the separation of two ion species.
$R$ is defined as

\begin{equation}
R(A;Z,\Delta_Z)=\frac{A_Z-A_{Z-\Delta Z}}{\sigma_{A_Z}+\sigma_{A_{Z-\Delta Z}}},
\label{eq4}
\end{equation}
where $A$ indicates the parameters used in the charge identification, $Z$ is the proton number and $\Delta Z$ is the proton number difference of two ion species.
For the normal distribution, two peaks can be resolved if $R \ge 2$, corresponding to less than $2.5\%$ of events from one peak to overlap with another peak.
In the case of $R=1$, two peaks can be identified with $\sim 16\%$ overlapping events.

\begin{figure}[h]\centering
	\includegraphics[angle=0,width=8cm]{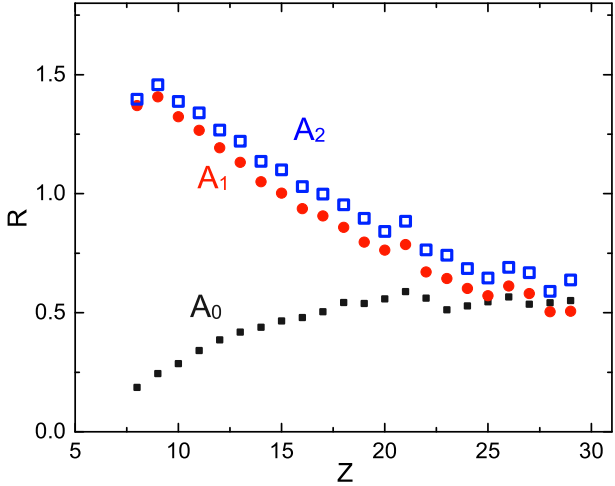}
	\caption{ (Colour online)
		For groups of ions with proton number difference $\Delta Z=2$, plot of charge resolution ratios $R$ by using $A_0$ (black solid block), $A_1$ (red solid dot) and $A_2$ (blue hollow block) as a function of proton number $Z$.
		\label{fig5}}
\end{figure}

\begin{figure}[h]\centering
	\includegraphics[angle=0,width=8cm]{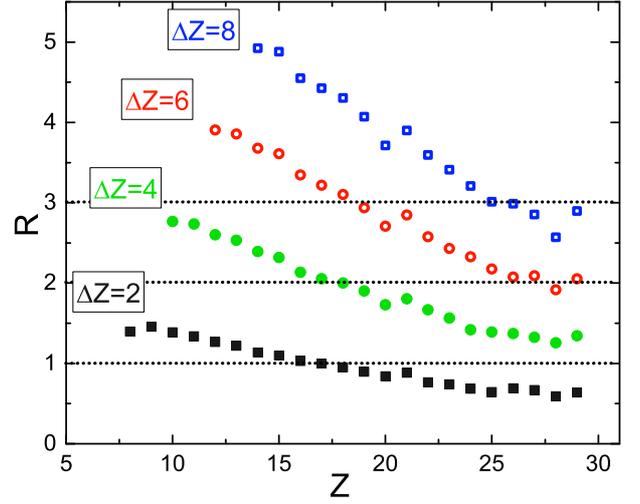}
	\caption{(Colour online)
		Using the constructed parameter $A_2$ in calculation, plot of charge resolution ratios $R$ as a function of proton number $Z$, for groups of ions with $\Delta Z=2$ (black solid block), $\Delta Z=4$ (green solid dot), $\Delta Z=6$ (red hollow dot),$\Delta Z=8$ (blue hollow block).
		\label{fig6}}
\end{figure}

Figure~\ref{fig5} shows the charge resolution ratios for different pairs of ions with $\Delta Z=2$.
The charge resolution ratios for $A_2$ ($R(A_2)$) are the largest for all ions, 
with an increase by $10\%-650\%$ as comparing with those for $A_0$.
The $A_2$ parameter provides the highest charge resolution.
Using $A_2$ values in the calculation, Figure~\ref{fig6} shows $R(A_2)$ for pairs of ions with various $\Delta Z$ as a function of $Z$.
The $R$ values are directly related to $\Delta Z/Z$.
In this experiment, pairs of ions with $\Delta Z/Z>0.22$ can clearly be resolved.

Only a moderate charge resolution was achieved in this experiment since the charge resolution was not specially optimized.
A better charge resolution can be realized with several technical improvements.
The charge resolution can be improved by extending the recording time for each injection, 
namely from the present 200~$\mu$s to 400-600~$\mu$s.
A lower number of stored ions in an individual injection will also lead to a higher resolution.
Two timing detectors, a new IMS method which is under development and will be applied in future experiments at CSRe \cite{Xu15(IMS), TOF, Xing, add3, Yan}, could also increase the charge resolution.
Last but not least, the MCP plates with 2~$\mu$m pore size will be adopted thus boosting the number of micro-channels.

\section{Summary}

In the IMS experiments, charge resolution provides crucial information for particle identification of ion pairs 
with $m/q$ ratios too similar to be resolved by their revolution times.
In an earlier work, the charge-resolved IMS was developed at the CSRe to resolve the $^{51}$Co$^{27+}$ and $^{34}$Ar$^{18+}$ ions.
However, the two ion types were not separated completely due to the limited charge resolution.
In this work a new analysis method is presented, which explicitly considers in the data analysis the average signal amplitudes, 
the detection efficiencies, and the number of stored ions.
The charge resolution of IMS is significantly improved for ions with $Z<22$, leading to a complete separation of $^{34}$Ar and $^{51}$Co.
The mass excess of $^{51}$Co is re-determined to be $-27332(41)$ keV, in agreement with the previous value \cite{11}.
The developed analysis method may be of a broader interest for storage-ring-based atomic and nuclear physics experiments using similar type of detectors.


\begin{thebibliography} {99}

\bibitem{1} M. Hausmann, F. Attallah, K. Beckert, {\it et al.}, First isochronous mass spectrometry at the experimental storage ring ESR, Nucl. Instr. and Meth. in Phys. Res. A, 446 (2000) 569. \href{https://doi.org/10.1016/S0168-9002(99)01192-4}{doi: 10.1016/S0168-9002(99)01192-4}

\bibitem{1_1} M. Hausmann, J. Stadlmann, F. Attallah, {\it et al.}, Isochronous Mass Measurements of Hot Exotic Nuclei, Hyperfine Interactions, 132 (2001) 291--297.
\href{https://doi.org/10.1023/A:1011911720453}{doi:10.1023/A:1011911720453 }

\bibitem{2_1} Yu. A. Litvinov, S. Bishop, K. Blaum, {\it et al.}, Nuclear physics experiments with ion storage rings, Nucl. Instr. and Meth. in Phys. Res. B, 317 (2013) 603--616. \href{https://doi.org/10.1016/j.nimb.2013.07.025}{doi:10.1016/j.nimb.2013.07.025 }

\bibitem{2_2} F. Bosch, Yu. A. Litvinov and Th. St{\"o}hlker, Nuclear physics with unstable ions at storage rings, Prog. Part. Nucl. Phys. 73 (2013) 84--140. \href{https://doi.org/10.1016/j.ppnp.2013.07.002}{doi:10.1016/j.ppnp.2013.07.002 }

\bibitem{2} Y. H. Zhang, Yu. A. Litvinov, T. Uesaka, H. S. Xu, Storage ring mass spectrometry for nuclear structure and astrophysics research, Phys. Scr., 91 (2016) 073002. \href{https://iopscience.iop.org/article/10.1088/0031-8949/91/7/073002}{doi:10.1088/0031-8949/91/7/073002 }

\bibitem{Steck} M. Steck and Yu. A. Litvinov, Heavy-ion storage rings and their use in precision experiments with highly charged ions, Prog. Part. Nucl. Phys., 115 (2020) 103811. \href{https://doi.org/10.1016/j.ppnp.2020.103811}{doi:10.1016/j.ppnp.2020.103811 }

\bibitem{3_1} B. Mei, H. S. Xu, Y. H. Zhang, {\it et al.}, Odd-even staggering in yields of neutron-deficient nuclei produced by projectile fragmentation, Phys. Rev. C, 94 (2016) 044615.
\href{https://journals.aps.org/prc/abstract/10.1103/PhysRevC.94.044615}{doi:10.1103/PhysRevC.94.044615}

\bibitem{3_2} C. W. Ma, Y. D. Song and H. L. Wei, Binding energies of near proton-drip line Z = 22-28 isotopes determined from measured isotopic cross section distributions, Sci. China-Phys. Mech. Astron., 62 (2019) 012013.
\href{https://doi.org/10.1007/s11433-018-9256-8}{doi:10.1007/s11433-018-9256-8}

\bibitem{3_3} Y. D. Song, H. L. Wei, C. W. Ma and J. H. Chen, Improved FRACS parameterizations for cross sections of isotopes near the proton drip line in projectile fragmentation reactions, Nucl.  Sci. Tech., 29 (2018) 96.
\href{https://doi.org/10.1007/s41365-018-0439-4}{doi:10.1007/s41365-018-0439-4}


\bibitem{8_1} J. Tr{\"o}tscher, K. Balog, H. Eickhoff, {\it et al.}, Mass measurements of exotic nuclei at the ESR, Nucl. Instr. and Meth. in Phys. Res. B, 70 (1992) 455--458. \href{https://doi.org/10.1016/0168-583X(92)95965-T}{doi:10.1016/10.1016/0168-583X(92)95965-T }

\bibitem{8} B. Mei, X. L. Tu, M. Wang, {\it et al.,} A high performance Time-of-Flight detector applied to isochronous mass measurement at CSRe, Nucl. Instr. and Meth. in Phys. Res. A, 624 (2010) 109. \href{https://doi.org/10.1016/j.nima.2010.09.001}{doi:10.1016/j.nima.2010.09.001 }

\bibitem{TOF} W. Zhang, X. L. Tu, M. Wang, {\it et al.,} Time-of-flight detectors with improved timing performance for isochronous mass measurements at the CSRe, Nucl. Instr. and Meth. in Phys. Res. A, 756 (2014) 1. \href{https://doi.org/10.1016/j.nima.2014.04.051}{doi:10.1016/j.nima.2014.04.051 }

\bibitem{8_2} T. Yamaguchi for the Rare RI Ring Collaboration, Present status of Rare-RI Ring facility at RIBF, Phys. Scripta T166 (2015) 014039. \href{https://iopscience.iop.org/article/10.1088/0031-8949/2015/T166/014039}{doi:10.1088/0031-8949/2015/T166/014039}

\bibitem{9} X. L. Tu, M. Wang, Yu. A. Litvinov, {\it et al.,} Precision isochronous mass measurements at the storage ring CSRe in Lanzhou, Nucl. Instr. and Meth. in Phys. Res. A, 654 (2011) 213. \href{https://doi.org/10.1016/j.nima.2011.07.018}{doi:10.1016/j.nima.2011.07.018}

\bibitem{10} Y. M. Xing, Y. H. Zhang, M. Wang, {\it et al.}, Particle identification and revolution time corrections for the isochronous mass spectrometry in storage rings, Nucl. Instr. and Meth. in Phys. Res. A, 941 (2019) 162331. \href{https://doi.org/10.1016/j.nima.2019.06.072}{doi:10.1016/j.nima.2019.06.072}

\bibitem{Liang} P. F. Liang, L. J. Sun, J. Lee,{\it et al.}, Simultaneous measurement of $\beta$-delayed proton and $\gamma$ emission of $^{26}$P for the $^{25}$Al($p,\gamma$), $^{26}$Si reaction rate, Physical Review C, Phys. Rev. C, 101 (2020) 024305. \href{https://link.aps.org/doi/10.1103/PhysRevC.101.024305}{doi:10.1103/PhysRevC.101.024305}

\bibitem{Sun} B. H. Sun, J. W. Zhao, X. H. Zhang {\it et al.}, Towards the full realization of the RIBLL2 beam line at the HIRFL-CSR complex, Science Bulletin, 63 (2018) 78. \href{https://doi.org/10.1016/j.scib.2017.12.005}{doi:10.1016/j.scib.2017.12.005}


\bibitem{r3} Y. Yamaguchi, M. Wakasugi, T. Uesaka, {\it et al.}, Construction of rare-RI ring at RIKEN RI Beam Factory, Nucl. Instr. and Meth. in Phys. Res. B, 317 (2013) 629--635. \href{https://doi.org/10.1016/j.nimb.2013.06.004}{doi:10.1016/j.nimb.2013.06.004}

\bibitem{yamaguchi} T. Yamaguchi, Y. Yamaguchi and A. Ozawa, The challenge of precision mass measurements of short-lived exotic nuclei: Development of a new storage ring mass spectrometry, Int. J. Mass Spectr. 349-350 (2013) 240--246. \href{https://doi.org/10.1016/j.ijms.2013.04.027}{doi:10.1016/j.ijms.2013.04.027}


\bibitem{CSRm} J. W. Xia, W. L. Zhan, B. W. Wei, {\it et al.}, The heavy ion cooler-storage-ring project (HIRFL-CSR) at Lanzhou, Nucl. Instr. and Meth. in Phys. Res. A, 488 (2002) 11--25. \href{https://doi.org/10.1016/S0168-9002(02)00475-8}{doi:10.1016/S0168-9002(02)00475-8}

\bibitem{1_5} B. Franzke, H. Geissel and G. M{\"u}nzenberg, Mass and lifetime measurements of exotic nuclei in storage rings, Mass Spectr. Rev., 27 (2008) 428--469. \href{https://onlinelibrary.wiley.com/doi/full/10.1002/mas.20173}{doi:10.1002/mas.20173}

\bibitem{Fu20} C. Y. Fu, Y. H. Zhang, M. Wang, {\it et al.}, Mass measurements for the $T_{z}=-2$ $fp$-shell nuclei $^{40}$Ti, $^{44}$Cr, $^{46}$Mn, $^{48}$Fe, $^{50}$Co, and $^{52}$Ni, Phys. Rev. C, 102 (2020) 054311. \href{https://journals.aps.org/prc/abstract/10.1103/PhysRevC.102.054311}{doi:10.1103/PhysRevC.102.054311}

\bibitem{3} X. Xu, M. Wang, K. Blaum,  {\it et al.}, Masses of neutron-rich $^{52\text{--}54}\mathrm{Sc}$ and $^{54,56}\mathrm{Ti}$ nuclides: The $N=32$ subshell closure in scandium, Phys. Rev. C, 99 (2019) 064303. \href{https://journals.aps.org/prc/abstract/10.1103/PhysRevC.99.064303}{doi:10.1103/PhysRevC.99.064303}

\bibitem{4} X. Xu, J. H. Liu, C. X. Yuan, {\it et al.}, Masses of ground and isomeric states of $^{101}\mathrm{In}$ and configuration-dependent shell evolution in odd-$A$ indium isotopes, Phys. Rev. C, 100 (2019) 051303(R). \href{https://journals.aps.org/prc/abstract/10.1103/PhysRevC.100.051303}{doi:10.1103/PhysRevC.100.051303}

\bibitem{5} Y. H. Zhang, P. Zhang, X. H. Zhou,  {\it et al.,} Isochronous mass measurements of ${T}_{z}=\ensuremath{-}1\phantom{\rule{4pt}{0ex}}fp$-shell nuclei from projectile fragmentation of $^{58}\mathrm{Ni}$, Phys. Rev. C, 98 (2018) 014319. \href{https://journals.aps.org/prc/abstract/10.1103/PhysRevC.98.014319}{doi:10.1103/PhysRevC.98.014319}

\bibitem{6} C. Y. Fu, Y. H. Zhang, X. H. Zhou,  {\it et al.,} Masses of the ${T}_{z}=\ensuremath{-}3/2$ nuclei $^{27}\mathrm{P}$ and $^{29}\mathrm{S}$, Phys. Rev. C, 98 (2018) 014315. \href{https://journals.aps.org/prc/abstract/10.1103/PhysRevC.98.014315}{doi:10.1103/PhysRevC.98.014315}

\bibitem{7} M. Z. Sun, X. H. Zhou, M. Wang, {\it et al.,} Precision mass measurements of short-lived nuclides at HIRFL-CSR in Lanzhou, Frontiers of Physics, Front. Phys., 13(6) (2018) 132112. \href{https://doi.org/10.1007/s11467-018-0844-5}{doi:10.1007/s11467-018-0844-5}

\bibitem{52Co} X. Xu, P. Zhang, P. Shuai, {\it et al.,} Identification of the lowest $T=2$, ${J}^{\ensuremath{\pi}}={0}^{+}$ isobaric analog state in $^{52}\mathrm{Co}$ and its impact on the understanding of $\ensuremath{\beta}$-decay properties of $^{52}\mathrm{Ni}$, Phys. Rev. Lett., 117 (2016) 182503. \href{https://journals.aps.org/prl/abstract/10.1103/PhysRevLett.117.182503}{doi:10.1103/PhysRevLett.117.182503}

\bibitem{Xu15} X. Xu, M. Wang, Y.H. Zhang, {\it et al.,} Direct mass measurements of neutron-rich $^{86}$Kr projectile fragments and the persistence of neutron magic number N=32 in Sc isotopes, Chin. Phys. C, 39 (2015) 104001. \href{https://doi.org/10.1088/1674-1137/39/10/104001}{doi:10.1088/1674-1137/39/10/104001}

\bibitem{9_1} X. Tu, H. Xu, M. Wang, {\it et al.}, Direct mass measurements of short-lived A=2Z-1 nuclides $^{63}$Ge, $^{65}$As, $^{67}$Se, and $^{71}$Kr and their impact on nucleosynthesis in the rp process, Phys. Rev. Lett. 106 (2011) 112501. \href{https://doi.org/10.1103/PhysRevLett.106.112501}{doi:10.1103/PhysRevLett.106.112501}

\bibitem{add1} P. Zhang, X. Xu, P. Shuai, {\it et al.}, High-precision $Q_{EC}$ values of superallowed 0$^{+}\to$0$^{+}$ $\beta$-emitters $^{46}$Cr, $^{50}$Fe and $^{54}$Ni, Phys. Lett. B, 767 (2017) 20--24. \href{https://doi.org/10.1016/j.physletb.2017.01.039}{doi:10.1016/j.physletb.2017.01.039}

\bibitem{add2} Y. Xing, K. Li, Y. H. Zhang, {\it et al.}, Mass measurements of neutron-deficient Y, Zr, and Nb isotopes and their impact on rp and $\nu$p nucleosynthesis processes, Phys. Lett. B, 781 (2018) 358--363. \href{https://doi.org/10.1016/j.physletb.2018.04.009}{doi:10.1016/j.physletb.2018.04.009}


\bibitem{11} P. Shuai, H. S. Xu, X. L. Tu, {\it et al.}, Charge and frequency resolved isochronous mass spectrometry and the mass of $^{51}$Co, Physics Letters B, 735 (2014) 327. \href{https://doi.org/10.1016/j.physletb.2014.06.046}{doi:10.1016/j.physletb.2014.06.046}

\bibitem{12} Y. H. Zhang, H. S. Xu, Yu. A. Litvinov, {\it et al.}, Mass Measurements of the Neutron-Deficient $^{41}$Ti, $^{45}$Cr, $^{49}$Fe, and $^{53}$Ni Nuclides: First Test of the Isobaric Multiplet Mass Equation in $fp$-Shell Nuclei, Phys. Rev. Lett., 109 (2012)  102501. \href{https://doi.org/10.1103/PhysRevLett.109.102501}{doi:10.1103/PhysRevLett.109.102501}

\bibitem{13} X. L. Yan, H. S. Xu, Yu. A. Litvinov, {\it et al.}, Mass measurement of $^{45}$Cr and its impact on the Ca-Sc cycle in X-ray bursts, Astrophys. J. Lett., 766 (2013) L8. \href{https://doi.org/10.1088/2041-8205/766/1/L8}{doi:10.1088/2041-8205/766/1/L8}

\bibitem{15} https://www.photonis.com/.


\bibitem{ps} W. Zhang, X. L. Tu, M. Wang, {\it et al.,} A timing detector with pulsed high-voltage power supply for mass measurements at CSRe, Nucl. Instr. and Meth. in Phys. Res. A, 755 (2014) 38. \href{https://doi.org/10.1016/j.nima.2014.04.031}{doi:10.1016/j.nima.2014.04.031}

\bibitem{Xu15(IMS)} X. Xu, M. Wang, P. Shuai, {\it et al.,} A data analysis method for isochronous mass spectrometry using two time-of-flight detectors at CSRe, Chin. Phys. C, 39 (2015) 106201. \href{https://doi.org/10.1088/1674-1137/39/10/106201}{doi:10.1088/1674-1137/39/10/106201}


\bibitem{Xing} Y. M. Xing, M. Wang, Y. H. Zhang, {\it et al.,} First isochronous mass measurements with two time-of-flight detectors at CSRe, Phys. Scr., T166 (2015) 014010. \href{https://doi.org/10.1088/0031-8949/2015/T166/014010}{doi:10.1088/0031-8949/2015/T166/014010}

\bibitem{add3} P. Shuai, X. Xu, Y. H. Zhang, {\it et al.}, An improvement of isochronous mass spectrometry: Velocity measurements using two time-of-flight detectors, Nucl. Instr. and Meth. in Phys. Res. B, 376 (2016) 311-315. \href{https://doi.org/10.1016/j.nimb.2016.02.006}{doi:10.1016/j.nimb.2016.02.006}


\bibitem{Yan} X. L. Yan, R. J. Chen, M. Wang, {\it et al.,} Characterization of a double Time-Of-Flight detector system for accurate velocity measurement in a storage ring using laser beams, Nucl. Instr. and Meth. in Phys. Res. A, 931 (2019) 52. \href{https://doi.org/10.1016/j.nima.2019.03.058}{doi:10.1016/j.nima.2019.03.058}


\end{thebibliography}
\end{document}